\documentclass[%
aip,
rsi,%
amsmath,amssymb,
reprint,%
]{revtex4-1}

\usepackage{graphicx}
\usepackage{dcolumn}
\usepackage{bm}

\begin{document}
	
	\title{Combining one and two photon polymerization for accelerated high performance (3+1)D photonic integration}
	
	\author{Adrià Grabulosa}
	\affiliation{FEMTO-ST Institute/Optics Department, CNRS \& University Bourgogne Franche-Comt\'e, \\15B avenue des Montboucons,
		Besan\c con Cedex, 25030, France}
	\author{Johnny Moughames}
	\email{Johnny.Moughames@femto-st.fr}
	\affiliation{FEMTO-ST Institute/Optics Department, CNRS \& University Bourgogne Franche-Comt\'e, \\15B avenue des Montboucons,
		Besan\c con Cedex, 25030, France}
	\author{X. Porte}%
	\affiliation{FEMTO-ST Institute/Optics Department, CNRS \& University Bourgogne Franche-Comt\'e, 
		\\15B avenue des Montboucons,
		Besan\c con Cedex, 25030, France
	}%
	\author{D. Brunner}
	\affiliation{FEMTO-ST Institute/Optics Department, CNRS \& University Bourgogne Franche-Comt\'e, \\15B avenue des Montboucons,
		Besan\c con Cedex, 25030, France
	}%

	\date{\today}
	
	\begin{abstract}
		
		Dense and efficient circuits with component sizes approaching the physical limit is the hallmark of high performance integration.
		Ultimately, these features and their pursuit enabled the multi-decade lasting exponential growth of components on integrated electronic chips according to Moore's law, which culminated with the high performance electronics we know today.
		However, current fabrication technology is mostly constraint to 2D lithography, and thermal energy dissipation induced by switching electronic signal lines presents a fundamental challenge for truly 3D electronic integration.
		Photonics reduces this problem, and 3D photonic integration is therefore a highly sought after technology that strongly gains in relevance due to the need for scalable application-specific integrated circuits for neural networks.
		Direct laser writing of a photoresin is a promising high-resolution tool for 3D photonic integration. 
		Here, we combine one and two-photon polymerization (TPP) for waveguide integration for the first time, dramatically accelerating the fabrication process and increasing optical confinement.
		3D additive printing is based on femtosecond TPP, while blanket irradiation with a UV lamp induces one-photon polymerization (OPP) throughout the entire 3D chip.
		We locally and dynamically adjust writing conditions to implement (3+1)D \emph{flash}-TPP: waveguide cores are printed with a small distance between neighboring writing voxels to ensure smooth interfaces, mechanical support structures are printed at maximal distance between the voxels to speed up the process.
		Finally, the entire chip's \emph{passive} volume not part of waveguide cores or mechanical support is polymerized in a single instance by UV blanket irradiation.
		This decouples fabrication time from the passive volume's size.
		We succeed in printing vertical single-mode waveguides of 6~mm length that reach up to NA = 0.16.
		Noteworthy, we achieve exceptionally low -0.26 dB injection and very low propagation losses of -1.36 dB/mm, which is within one order of magnitude of standard integrated silicon photonics.
		Finally, the optical performance of our waveguides does not deteriorate during 120 days within our measurement uncertainty.
		
	\end{abstract}
	
	\maketitle

	\section{Introduction} 
	
	Additive manufacturing based on 3D printing is an innovative tool to create complex 3D devices.
	Direct laser writing (DLW) associated with two-photon polymerization (TPP) allows the creation of micron to sub-micrometer three-dimensional (3D) structures in diverse fields, such as micromechanical systems \cite{Maruo2003}, microrobotics components \cite{Ji2021} or biosciences \cite{Pitts2000}. 
	In DLW, a tightly focused laser spot is translated through a photoresin to form a solid inside the TPP process's reactive volume, which results in a sub-micron voxel \cite{Hong2004}.
	In photonics, the DLW-TPP technique has been used to fabricate free-form and transformational components \cite{Ergin2010,Ristok2020,Dietrich2018}, point-to-point photonic wire-bondings \cite{Lindenmann2012}, waveguides \cite{Porte2021}, spatial-filters \cite{Moughames2020}, graded-index lenses \cite{Moughames2016} and photonic components \cite{Buckmann2012,Lin2020}.
	Simultaneously, this technique is advantageous for integrated photonic circuits \cite{Gissibl2016,Ulas2020} due to its ability to locally and dynamically modify optical properties on feature sizes below the Abbe resolution limit.
	
	\begin{figure}[h!]
		\centering
		\includegraphics[width=0.82\linewidth]{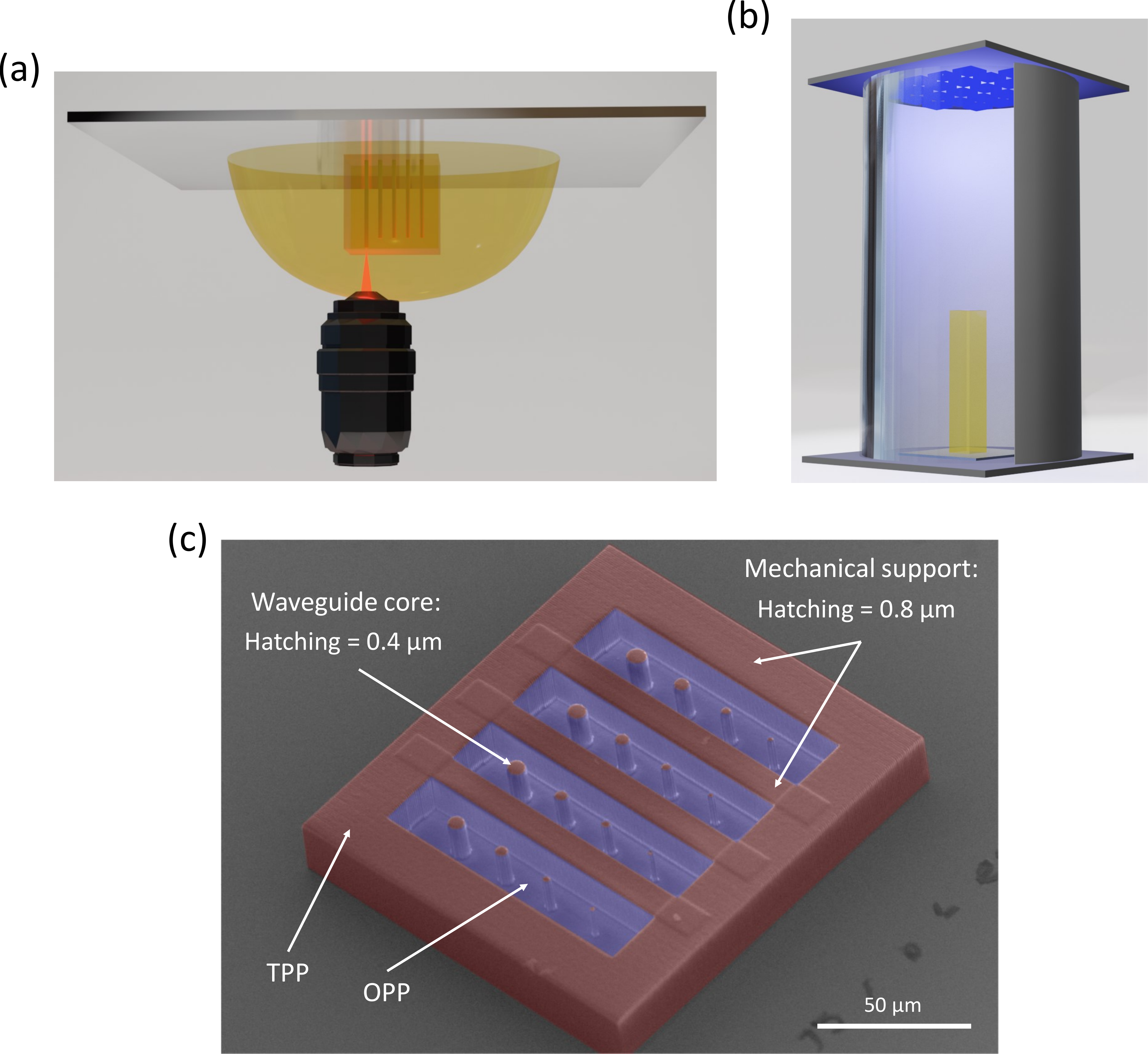}
		\caption{\emph{Flash}-TPP for 3D integrated photonics. (a) Illustration of the dip-in process for direct laser writing of 3D waveguides. The IP-S resin is polymerized via TPP using a 780 nm laser and a 25X microscope objective. (b) A UV light source polymerizes the unexposed regions of the structure. The \emph{flash}-TPP concept leverages both, one and two photon polymerization. (c) SEM micrograph of a 3D-printed cross-section cutting through a cuboid integrating 16 waveguides with diameters ranging from $d\in \{0.8:0.4:6.8 \}~\mu$m. The waveguide cores are printed in high-resolution for maximizing optical performance, while structures only serving mechanical stability are printed with the lowest possible spatial resolution. Red colour represents the TPP-printed regions, blue colour represents regions that are exclusively polymerized with one-photon polymerization (OPP) via UV exposure.     
		}
		\label{fig:Fig1}
	\end{figure}
	
	Photonic waveguides are of fundamental importance for many optical components and to realize their implementation in day-to-day technologies \cite{Lindenmann2012,Kumar2018}.
	The capability of additively integrating photonic circuits with micron feature sizes opens new perspectives for applications, such as the scalability of integrated photonic neural network circuits \cite{Moughames2020,Moughames2021}, chip-to-chip interconnects \cite{Nesic2019} or in building blocks for photonic waveguide integration in 3D \cite{Pyo2016,Brunner2020}.
	Micromachining of photonic waveguides by DLW-TPP is currently explored in diverse materials, i.e. glasses, crystals and polymers \cite{Valle2009}. 
	For DLW additive manufacturing, the working distance of the microscope objective does not constraint the fabrication process and allows single-step printing on different substrates.
	The technique is therefore an excellent candidate for manufacturing complex and large integrated photonic circuits directly on top of other, e.g. semiconductor chips.
	
	In standard photonic waveguides, light confinement is caused by total internal reflection at the interface between core and cladding. 
	The main condition for this is that the refractive index of the core $n_{\text{core}}$ is larger than the one of the cladding $n_{\text{cladding}}$.
	For polymer waveguides in which the surrounding media is air, the core-cladding refractive index difference $\Delta{n}=n_{\text{core}}-n_{\text{cladding}}\approx0.5$, is large and easily leads to multi-mode propagation at the feature sizes achievable with TPP for visible or near infrared (NIR) wavelengths \cite{Snyder}. 
	Single-mode propagation can be ensured by a core diameter below the single-mode cut-off diameter.
	For $\Delta{n}\approx0.5$ this is out of scope for standard and single-step additive DLW techniques.
	However, single-mode waveguides are a fundamentally important and are broadly employed in integrated photonics, low-loss fibers \cite{polym12112485} or for quantum and optical communications \cite{Sartison2021,Nature2021}.
	To overcome this limitation, we have recently demonstrated a second approach where we leveraged precise control of the local degree of TPP polymerization and in turn over $\Delta$n on the level required for single-mode waveguides with $\mu$m core diameter \cite{Porte2021}. 
	
	In this article, we significantly advance our (3+1)D fabrication methodology, here based on the commercially available IP-S photoresist \cite{Schmid2019}. 
	In order to accelerate fabrication while simultaneously improving waveguiding performance, we combine three concepts dedicated to fabricating the three essential parts of a circuit: (i) the waveguide cores, (ii) waveguide cladding and (iii) mechanical support ensuring the stability of a integrated circuit.
	Waveguide cores require $\mu$m resolution and smooth interfaces with their cladding to ensure low-loss propagation. 
	We therefore use TPP with a fine resolution in the $(x,y)$-plane, i.e. a small hatching distance $h$, and a carefully optimized TPP writing power. 
	Areas only acting as mechanical support do not require smooth surfaces and we use large TPP-writing powers and low hatching resolution in order to reduce fabrication time. 
	Finally, the majority of an integrated photonic circuit comprises material with an uniform refractive index lower than the one of waveguide cores. 
	This, we achieve in a single-shot for the entire circuit via blanket ultraviolet (UV) irradiation with a carefully controlled exposure dosage to ensure optical confinement to the waveguide cores and single-mode propagation. 
	We achieve low propagation losses of -1.36 dB/mm and excellent mechanical stability enabling structures that approach centimeter scales, all while reducing fabrication times by $\approx$ 90$\%$ compared to full TPP fabrication of the same circuit with a constant hatching distance. 
	
	\section{Accelerate 3D waveguide printing with \emph{flash}-TPP} 
	
	One-photon polymerization (OPP) is often employed to process thin material layers such as in standard semiconductor photo-lithography. 
	The process is based on polymerizing a photosensitive resin using blue or UV light. 
	The surface of the thin resin-layer is scanned or irradiated through a photomask, and the liquid resin solidifies in the light-exposed regions. 
	Repeating this process layer-by-layer enables the fabrication of 3D structures.
	However, to achieve complex and truly 3D-circuits this process requires a large number of photo masks and results in challenges like intra-layer alignment.  
	Unlike OPP, the photocurable resin that is transparent for a NIR femtosecond (fs) laser allows directly printing deep inside the resin's volume via TPP.
	The basic principle of TPP relies on transforming monomer molecules located within the femtosecond writing laser's focus into macromolecules in a solid state.
	For the commercially available IP family thermosetting photoresists, i.e. IP-S, IP-Dip, IP-L and IP-G, the resin is essentially transparent from $\sim$633 nm until $2400~$nm.
	
	The Nanoscribe GmhH (Photonic Professional GT) TPP system used for printing our structures is equipped with a fs-laser operating at 780 nm, and galvanometer mirrors for rapid beam movement in the lateral directions. 
	The femtosecond laser is tightly focused into the resin through an objective lens of high numerical aperture (25X magnification and NA $=0.8$). 
	TPP leads to the formation of a rugby-ball-shaped polymer within the focal spot, the writing voxel, and the monomer resin in its surrounding remains unmodified. 
	The voxel-size depends mainly on the laser exposure dose, which is linked to the laser power and the scanning speed \cite{Sun2003}. 
	This voxel-size is relevant when considering feature-sizes of few microns due to a diffusion process that leads to a local modification of refractive index gradients of the polymerized structures after development.  
	In the TPP lithography process, the liquid negative-tone IP-S photoresist, with $n\approx$ 1.51 when fully TPP polymerized \cite{Gissibl2017,Li2019}, acts as an immersion medium for the objective lens. 
	In particular, the polymerization mechanism in the IP-S photoresist is catalyzed via a photo-initiator based on aromatic ketones, which accelerates the polymerization process by creating reactive species, i.e. free radicals, cations or anions \cite{Photo}.
	After polymerization, the unexposed photoresist is removed in a two-step development process using propylene-glycol-methyl-ether-acetate (PGMEA) as a developer for 20 minutes (min), followed by rinsing in isopropyl alcohol (2-propanol) for 3-5 min.
	
	One challenge of 3D TPP following the classical approach is that printing circuits with mm$^3$ volumes quickly results in unfeasible fabrication time in excess of 20h. 
	However, 3D waveguide circuits are defined via the trajectories of waveguide cores, while waveguide claddings are inherently realized through the surrounding, 'unstructured' volume of a lower refractive index. 
	One can therefore constrain the high-resolution and slow TPP process to creating the waveguide cores, and use indiscriminate, hence fast single-step blanket exposure above the resin photo-initiator's absorption energy to develop the entire remaining chip in one \emph{flash}.
	The working principle of \emph{flash}-TPP is illustrated in Fig. \ref{fig:Fig1}(a-c). 
	Figure \ref{fig:Fig1}(a) depicts the usual dip-in TPP printing procedure, where the microscope objective is directly immersed into the resin. 
	The printed structure is then developed using the previously described steps. 
	Next, we transfer the developed circuit to a UV chamber (Rolence Enterprise Inc., LQ-Box model, 405 nm wavelength, 150 mW/cm$^2$ average light intensity) as shown in Fig. \ref{fig:Fig1}(b), and we control the OPP dosage $D$ via the duration of the UV exposure. 
	Here, we characterize the effects of exposure times between 0 and 60 s.
	
	We further accelerate the fabrication process by locally and dynamically adjusting TPP writing conditions according to the functionality of each circuit's component. 
	Minimizing propagation losses requires minimizing the roughness at the core-cladding interface and we use a small spacing between voxels in the (x,y)-plane, i.e. with a small hatching distance $h$. 
	At the same time, we can maximize the vertical distance between consecutive slices, or slicing distance $s$, which we found not to significantly affect the core-cladding interface roughness and equally accelerates the printing process.
	Crucially, the writing power needs to be adjusted in order not to overexpose the TPP-written voxels, which results in micro-explosions and high propagation losses. 
	Other sections, such as the outer cladding of a 3D chip or internal columns for support do not interact with the optical signals, and hence they simply need to be mechanically sturdy. 
	We therefore maximally increase the hatching distance, and since printing time scales $\propto (h^{2}s)^{-1}$ this has a significant impact. 
	The structure in Fig. \ref{fig:Fig1}(c) shows an scanning electron microscopy (SEM) micrograph that illustrates the overall concept. 
	A circuit is first fabricated in a single 3D direct laser TPP lithography (red region) step, which is followed by OPP via irradiation with the UV light source (blue regions) for several seconds.

	\section{Two-photon polymerization fabrication parameters} 
	
\begin{figure}[h!]
		\centering
		\includegraphics[width=0.70\linewidth]{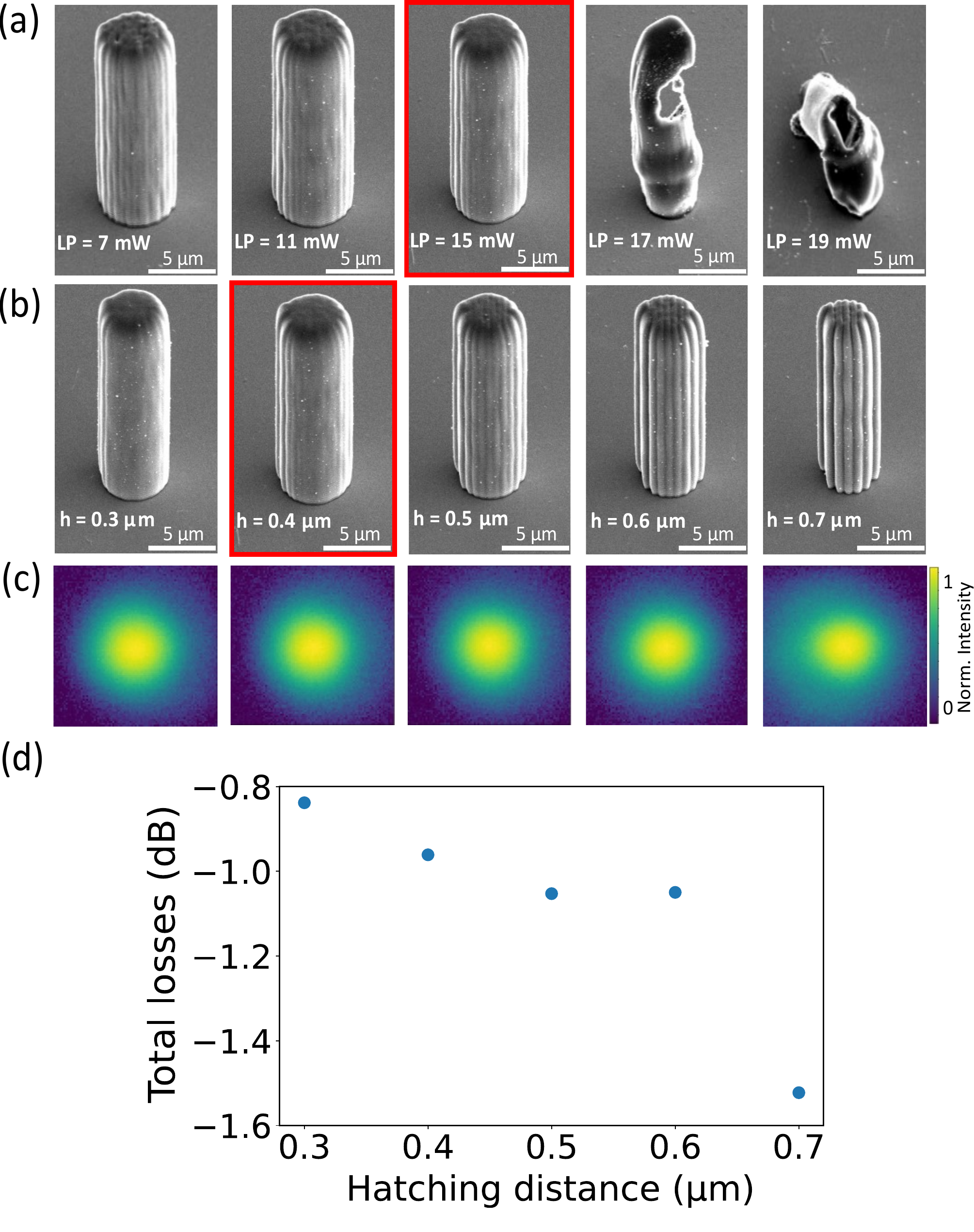}
		\caption{Structural and optical characterization of the TPP-printed waveguide cores. (a) SEM micrographs for TPP laser power LP $\in \{7\mathbin{,}\dots\mathbin{,}19 \}$~mW of $5~\mu$m diameter and $20~\mu$m heigh waveguides. Hatching $h$ and slicing $s$ distance are fixed to $0.4~\mu$m and $1~\mu$m, respectively. Best combination highlighted in red. (b) Impact of hatching distance $h\in \{0.3:0.1:0.7\}~\mu$m. Slicing distance and LP fixed to $1~\mu$m and 15 mW, respectively. Best combination highlighted in red. (c) Output intensities of a $\tilde{d}=3.3~\mu$m diameter and $300~\mu$m heigh waveguide fabricated with the same hatching, slicing and LP parameter parameters as in (b). Sample cured with UV light during 20s. Only for the largest hatching distance one can see weaker optical confinement and scattering of the optical mode. (d) Total losses versus hatching distance $h$ for the same waveguides as in (c). 
		}
		\label{fig:Fig2}
	\end{figure}

	We printed a set of five free-standing waveguide cores with $20~\mu$m height and $d=5~\mu$m diameter using a range of TPP laser powers (LP) and a constant hatching distance of $h=0.4~\mu$m on a fused silica substrate. 
	As globally fixed parameters in all our fabrications reported in this article we use a scanning speed of 10~mm/s and a slicing distance of $s=1~\mu$m.
	A SEM micrograph after development is shown in Fig. \ref{fig:Fig2}(a), with LP $\in \{7, \dots, 19 \}$~mW. 
	Waveguide cores shown in the first two images were printed with LP $=7~$mW and LP $=11~$mW, which results in rough and inhomogeneous surfaces. 
	Increasing the laser power to 15 mW leads to larger TPP voxels and consequently a smoother surface. 
	In contrast, exceeding LP $=15~$mW leads to overpolymerized waveguides and burning of the polymer, see the last two images of Fig. \ref{fig:Fig2}(a).
	We therefore select LP $=15~$mW and proceed to optimize the second parameter by scanning the hatching distance from $h\in \{0.3:0.1:0.7\}~\mu$m, with results shown in the SEM micrographs of Fig. \ref{fig:Fig2}(b).
	The first two micrographs, with respectively $h=0.3~\mu$m and $h=0.4~\mu$m, reveal smooth surfaces compared to the last three.
	We therefore select $h=0.4~\mu$m for the following investigations because, first, the visual inspection of the surface quality shows only a negligible difference compared to $h=0.3~\mu$m, but using $h=0.4~\mu$m reduces the fabrication time by a factor 1.8. 
	Second, for $h=0.3~\mu$m results were not always reproducible, a finding we attribute to spontaneous micro-burnings within the waveguide's cores as the smaller hatching distance increases the accumulative TPP-irradiation.
	SEM analysis showed that for these parameters the diameter of waveguide cores exceed the design-diameter $d$ by $d_0 = 0.5~\mu$m due to the non-negligible TPP voxel-size. 
	Hence, for the following investigations we consider the effective diameter $\tilde{d} = d + d_{0}$.
	
	Finally, to carry out an initial evaluation of the propagation losses on the hatching distance $h$, we have characterized five printed waveguides with length of $300~\mu$m at $\lambda_0=660$ nm wavelength with identical parameters as in Fig. \ref{fig:Fig2}(b). 
	Figure \ref{fig:Fig2}(c) shows the output intensities of the waveguides printed with $h\in \{0.3:0.1:0.7\}~\mu$m at a constant diameter of $\tilde{d}=3.3~\mu$m with LP $=15~$mW and $s=1~\mu$m. 
	After development the samples were UV cured during 20s. 
	The first four outputs have practically identical intensity distribution, and a notable reduction in optical confinement can only be found for $h=0.7~\mu$m. 
	This can also be seen in Fig. \ref{fig:Fig2}(d), which characterizes overall optical losses (injection, propagation, outcoupling) on $h$.
	Overall optical losses slowly grow with $h$ until they dramatically increase for $h=0.7~\mu$m.
	We consequently select $h=0.4~\mu$m and LP $=15~$mW for the fabrication of the TPP-printed waveguide cores, mechanical support structures were printed with low-resolution ($h=0.8~\mu$m) to reduce printing times. 
	We introduced walls between rows of waveguides in order to maintain a flat top-surface of the cuboids to counteract the effect of shrinkage during sample development, which otherwise results in buckling of the waveguides.

	\section{Modal confinement}
	
	We 3D-printed cuboids embedding 16 waveguides with diameters ranging from $\tilde{d}\in \{1.3:0.4:7.3 \}~\mu$m and height of $300~\mu$m. 
	We use UV exposure times for the OPP of 0, 5, 20 and 60s, which results in irradiation doses $D$ of 0, 750, 3000 and 9000 mJ/cm$^2$, respectively.
	
	\begin{figure}[h!]
		\centering
		\includegraphics[width=0.95\linewidth]{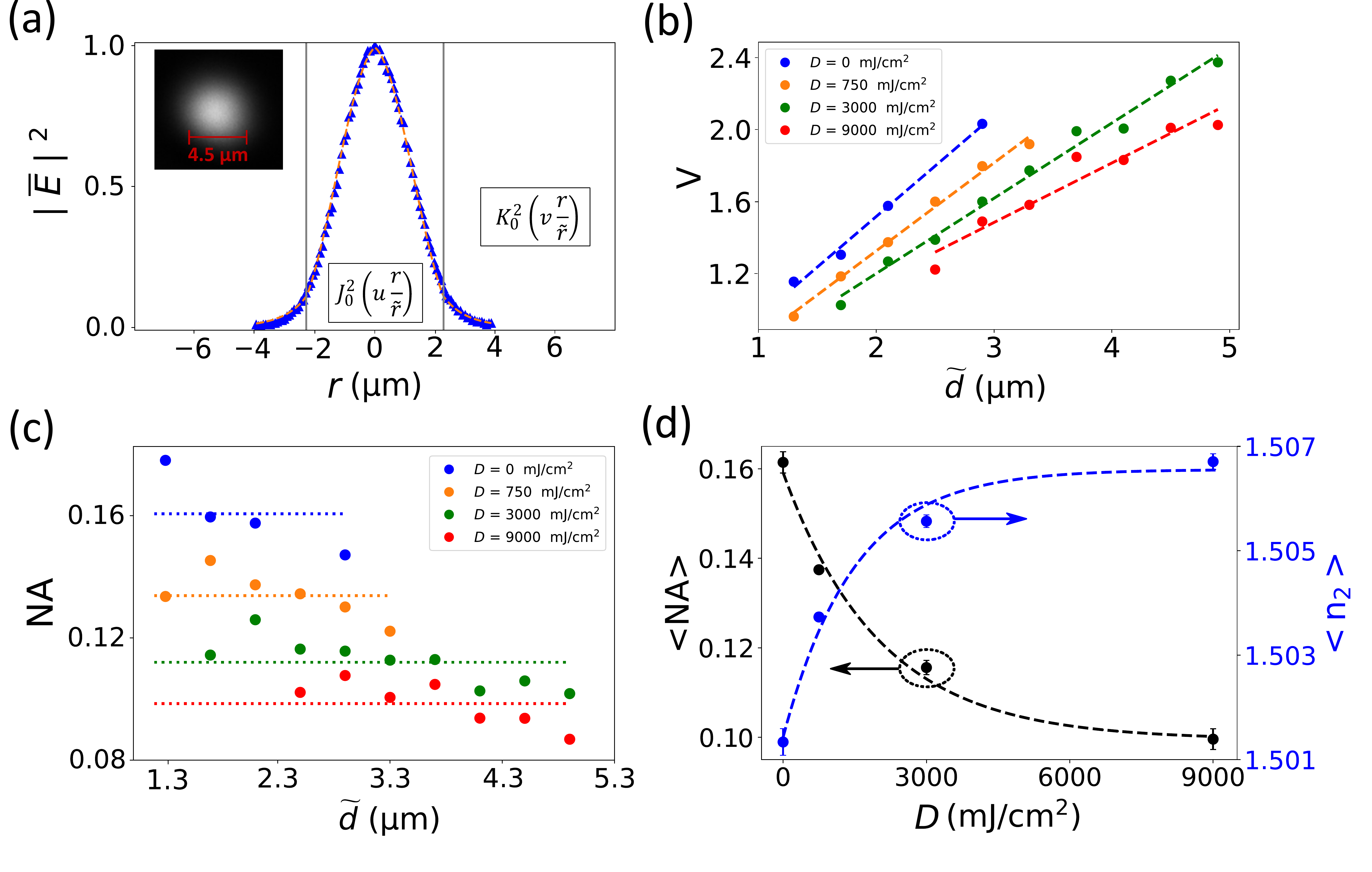}
		\caption{Modal confinement of 3D-printed waveguides cured via blanket UV exposure. (a) Output intensities (blue triangles) and fundamental LP$_{01}$ mode fit (dashed orange lines) of $\tilde{d} = 4.5~\mu$m waveguide cured with $D$ = 3000 mJ/cm$^2$. (b) Normalized frequency $V$ versus effective diameter $\tilde{d}$, OPP curing with 0 (blue dots), 750 (orange), 3000 (green) and 9000 (red) mJ/cm$^2$. Dashed lines indicate the linear regression of the experimental data. (c) Numerical aperture NA values versus effective waveguide $\tilde{d}$ dependence. Dashed lines represent the average <NA> for each UV dose $D$. (d) <NA> (black) and $<n_2>$ (blue) versus UV dose. Dashed lines indicate the exponential fit to experimental data.
		}
		\label{fig:Fig3}
	\end{figure}
	
	By fitting the experimental output intensities for different diameters, we extract the waveguide's normalized frequency $V = \frac{\pi}{\lambda_0}\tilde{d}~$NA, where NA $= \sqrt{n_1^2-n_2^2}$ is the numerical aperture and $n_1\approx$ 1.51 and $n_2$ are the refractive indices of core and cladding, respectively. 
	Each waveguide output intensity is fitted to the fundamental LP$_{01}$ mode for diameters below the cut-off condition for a second propagating mode, as shown in Fig. \ref{fig:Fig3}(a) for an exemplary waveguide of $\tilde{d}=4.5~\mu$m that was UV cured with $D$ = 3000 mJ/cm$^2$. 
	The LP$_{01}$ mode intensity profile in step-index (STIN) waveguides is described by $J_0^2$($u$$\frac{r}{\widetilde{r}})$ for $\lvert r\rvert$ < $\widetilde{r}$ and 
	$K_0^2$($v$$\frac{r}{\tilde{r}})$ for $\lvert r\rvert$ > $\tilde{r}$, where $\tilde{r}=\tilde{d} / 2$ and $u$, $v$ the modal parameters \cite{Snyder}. 
	Therefore, we individually determine the normalized frequency $V = \sqrt{u^2+v^2}$ for each waveguide and Fig. \ref{fig:Fig3}(b) shows the experimental characterization for all effective waveguides diameters $\tilde{d}$ cured using different UV doses $D$. 
	It is clear that the shorter we expose the circuit, the larger the normalized frequency. 
	From the slope of the linear regressions (dashed lines) in Fig. \ref{fig:Fig3}(b) we obtain the average NA for each UV exposure, which is shown in Fig. \ref{fig:Fig3}(c). 
	
	As it can be seen in Fig. \ref{fig:Fig3}(c), the NA appears to slightly decrease for larger diameters. 
	We attribute this to the diffusion during development, which leads to a smoother refractive index transition between waveguide core and cladding and consequently converts the STIN profile to an effective gradient index (GRIN) profile \cite{Ocier2020,TuingZukauskas2015}. 
	This phenomenon would additionally increase the NA for small diameter waveguides.
	
	Finally, Fig. \ref{fig:Fig3}(d) shows the evolution of the average numerical aperture <NA> and refractive index of the cladding $<n_2>$ on the UV exposure dose. 
	The excellent agreement with the exponential fit demonstrates that we can precisely tune the numerical aperture of the different waveguides, which decreases exponentially till it reaches a plateau. 
	Therefore, considering the refractive index of the core ($n_1$) is approximately constant once fully TPP-polymerized \cite{Schmid2019}, all variations of NA versus exposure dose $D$ can be assigned to the refractive index of the cladding following $n_2=\sqrt{{n_1}^2-{\textrm{NA}}^2}\exp{(\kappa·D)}$, by which we obtain the OPP factor of the IP-S photoresist $\kappa = 6.7\cdot10^{-4}$ mJ$^{-1}$cm$^2$ from the exponential fit.

	\section{Propagation losses}
	
	To evaluate losses we fabricated a set of waveguides with lengths spanning from 0.1 to 6 mm.  The diameter is $\tilde{d}=3.7~\mu$m and the UV expose dose $D$ = 3000~mJ/cm$^2$, which provide high NA while remaining single-mode. 
	Minimizing injection losses requires adiabatically modifying the waveguide core diameter to transition the mode size of the free-space input to the one of the waveguide.  
	The angle of such tapers, i.e. the taper-rate, needs to be sufficiently small to not excite higher order modes and hence to satisfy the adiabatic transition.  
	We minimized the optical losses of a set of tapers with lengths $\in\{10:10:50\}~\mu$m and waveguide core diameters $\tilde{d}\in \{3.7:0.4:5.7\}~\mu$m in order to determine the best taper-rate for efficient mode coupling.  
	These we found to be tapering from an taper-input diameter of $4.9~\mu$m to the target waveguide diameter of $3.7~\mu$m during a taper-length of $40~\mu$m.  
	
\begin{figure}[h!]
		\centering
		\includegraphics[width=0.85\linewidth]{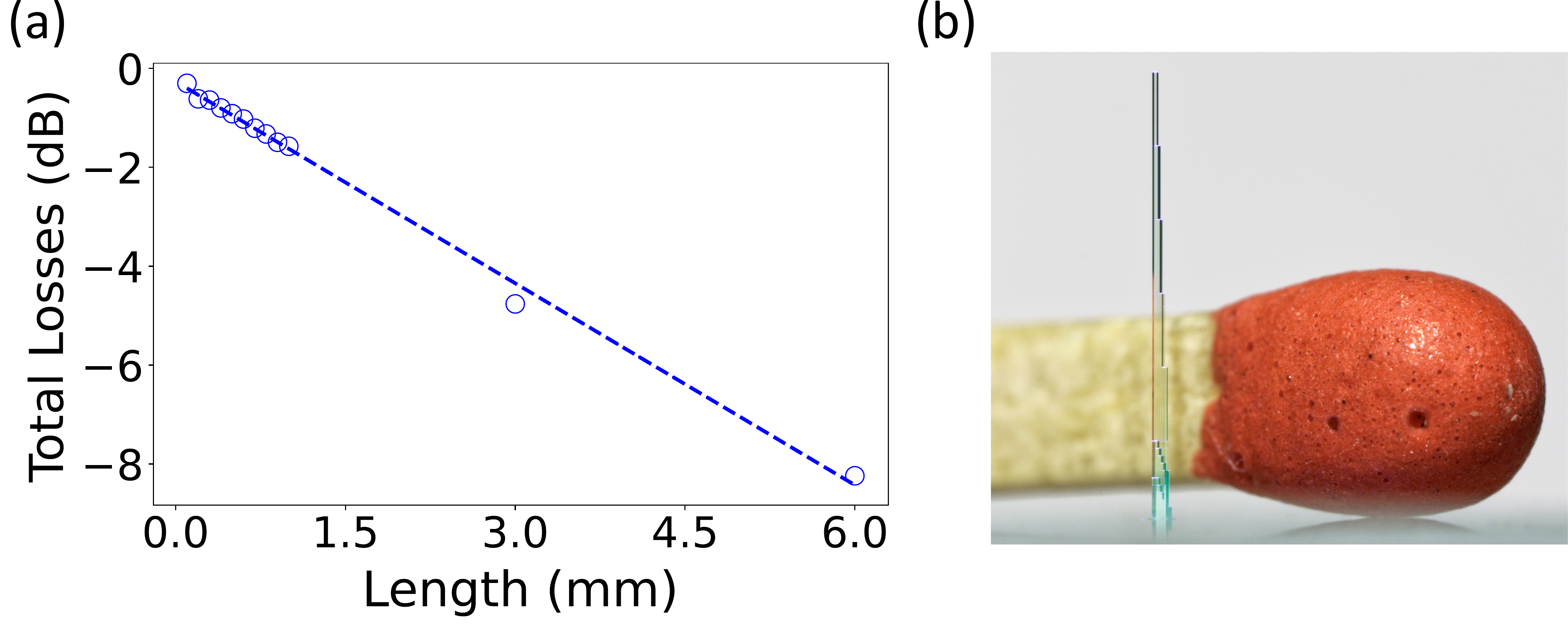}
		\caption{(a) Injection and propagation losses of the fundamental LP$_{01}$ mode for $\tilde{d}=3.7~\mu$m and $D$ = 3000 mJ/cm$^2$ UV dose. We find -0.26~dB injection and -1.36 dB/mm propagation losses. (b) Photography of the structure integrating the 6 mm long waveguide size-scaled to a match.   
		}
		\label{fig:Fig4}
	\end{figure}

	Figure \ref{fig:Fig4}(a) depicts the total losses for the fundamental LP$_{01}$ mode on a semi-logarithmic scale, and via the linear fit we determine -1.36 dB/mm propagation and -0.26 dB injection losses. 
	Our measurements are highly reproducible across the wide range of propagation lengths we here examined, in particular considering that each waveguide longer than 0.5 mm was fabricated on a different substrates.
	This demonstrates the excellent reproducibility and high quality of our (3+1)D \emph{flash}-TPP fabrication as well as of our optical characterization.
	
	Propagation losses do not yet reach the bulk absorption of the IP-S photoresist, which is $\approx$ 0.055 dB/mm \cite{Schmid2019}, and for now are around one order of magnitude above the current state-of-the-art for standard silicon photonic waveguides \cite{Vlasov2004}.
	Compared to single-pass DLW of waveguides in fused silica glass, our propagation losses are comparable at our characterization wavelength, while we achieve more than an order of magnitude lower injection losses \cite{Amorim2019}.
	Still, our results represent a factor 5 improvement compared to previously 3D-printed waveguides \cite{Moughames2021}.
	Figure \ref{fig:Fig4}(b) shows the 6 mm long waveguide printed for this study, imaged next to a match for scale comparison. 
	
	\section{Temporal stability}
	
	As a final study, we investigated aging of the 3D-printed waveguides, as temporal stability naturally is of vital importance for reliability.
	Aging could originate from such effects as thermal diffusion \cite{Zhang2006}, shrinkage \cite{Ovsianikov2009} or undesired photo-polymerization of areas with low degree of polymerization \cite{Schmid2019}. 
	
\begin{figure}[h!]
		\centering
		\includegraphics[width=0.65\linewidth]{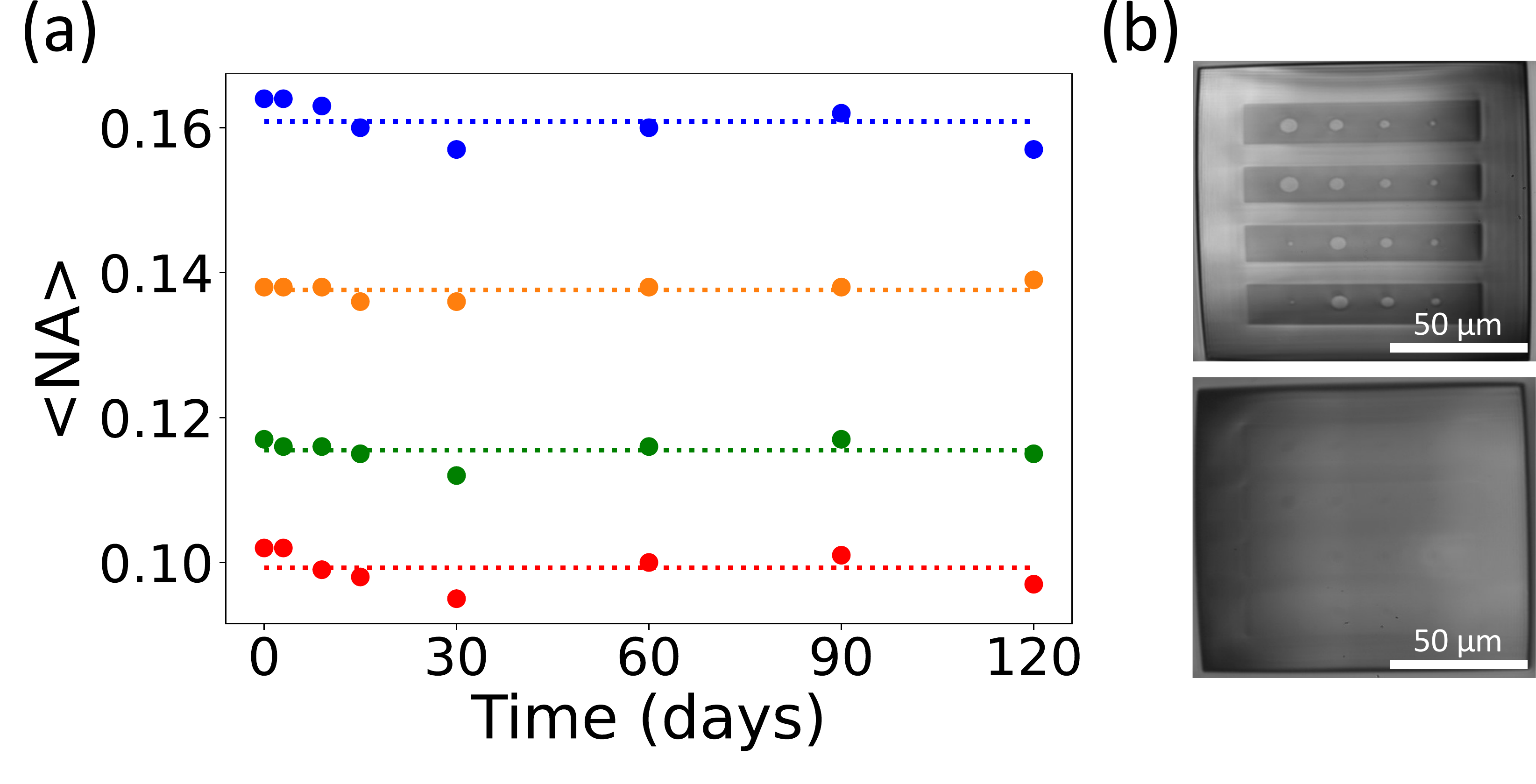}
		\caption{Temporal stability. (a) Average numerical aperture $<$NA$>$ over time for 3D-printed waveguides cured via blanket UV exposure. Dashed lines represent the average values for the different UV doses, showing no significant effect of aging. (b) Optical microscope image showing the effect of direct sun exposure for $\sim$15 days during summer in middle Europe, before (upper-panel) and after (bottom-panel) exposure.  
		}
		\label{fig:Fig5}
	\end{figure}

	The \emph{flash}-TPP concept relies on incomplete polymerization of the areas corresponding to waveguide cladding, hence some photo-initiators could remain after development. 
	If unintended post-development optical exposure is sufficient to initiate the photo-initiator's reactions, the waveguide cladding could in principle be polymerized towards a higher degree, i.e. increasing its refractive index. 
	Consequently, $\Delta$n and the related optical properties would be altered in time. 
	To explore the temporal stability we measured the waveguide's NA over time. 
	All structures have been kept under standard irradiation intensity of our lab and have not been shielded from room light. 
	As depicted in Fig. \ref{fig:Fig5}(a), the NA of all waveguides explored in the previous sections do not suffer any relevant change during their continuous characterization spanning 120 days. 
	
	Finally, we also evaluated the waveguide's resilience to direct sun exposure, for which we placed one sample on a window sill during summer months in central Europe.
	This resulted in the almost complete polymerization of the cladding area after $\sim$15 days and hence the erasure of the light-guiding structures, which can be seen from the microscope image in Fig. \ref{fig:Fig5}(b). 
	The direct and unprotected UV radiation from the sun therefore appears to be sufficiently strong \cite{Kerr2008} to launch the photo-initiators reaction and thus the polymerization process.
	We also examined the effect of direct sun exposure on waveguides for which the cladding was printed with a low TPP power dose (i.e. 1, 2 and 3 mW), and no difference compared to the \emph{flash}-TPP was found; the structures equally got erased by the direct sun exposure. 
	Therefore, we can conclude that the aging under direct exposure to sun light is through the general augmentation in polymerization-degree and does not depend on the polymerization technique used during fabrication. 
	
	\section{\emph{Flash}-TPP printing time}
	
	Printing time is generally proportional to a circuit's volume, and the relative time that can be saved due to \emph{flash}-TPP depends on the ratio between the areas requiring TPP and OPP exposure.
	Here we assume that this ratio remains constant along a circuit's z-positions.
	In previous work on (3+1)D printing we found that waveguides with a radius of $r\approx2.5~\mu$m need to be separated by $l\approx 6~\mu$m of cladding in order to have negligible interaction \cite{Porte2021}.
	Cross-talk essentially depends on the relationship of confinement and separation, and scaling of $l$ and $r$ should therefore remain comparable for integrated single-mode circuits with not too large $\Delta$n.
	Using TPP to print the entire structure requires a fabrication time $T_{\textrm{TPP}}\propto (l+2r)^2$, while \emph{flash}-TPP requires $T_{\textrm{flash}}\propto \pi{r}^2$.
	As a result, the relative duration $\Gamma$ of \emph{flash}-TPP relative to classic TPP is given by
	\begin{equation}\label{eq:TimeSave}
		\Gamma = \frac{\pi r^2}{(l+2r)^2} = \frac{\pi}{4}\frac{1}{(1+l/2r)^2},
	\end{equation}
	\noindent which is also the filling factor of waveguide cores within a 3D integrated photonic chip.
	Therefore, even for a circuit volume with maximal density of waveguide cores, the usual objective of circuit integration, using $l$ and $r$ defined above the fabrication time is reduced to $\Gamma\approx16\%$ compared to using TPP alone.
	This agrees with our experience, where using \emph{flash}-TPP reduces the printing time to only $\Gamma\approx10\%$ compared to classical TPP. As a clear example, printing our 6 mm macroscopic structure exclusively using TPP only takes $\sim$24 hours, whereas this decreased to $\sim$3 hours ($\Gamma=12\%$) using \emph{flash}-TPP.

	\section{Conclusion}
	
	We have developed a simple and fast (3+1)D lithography configuration called \emph{flash}-TPP. 
	This novel manufacturing methodology is based on combining DLW-TPP and OPP, here demonstrated with the commercially available IP-S photoresist, for the fabrication of polymer-cladded single-mode 3D optical waveguides.
	The \emph{flash}-TPP concept synergistically combines three princples.  
	First, waveguide cores are printed via DLW-TPP using a precisely calibrated TPP laser power and with high horizontal resolution by minimizing the hatching distances between voxels ($h=0.4~\mu$m),  ensuring low propagation losses due to smooth interfaces. 
	Second, areas purely serving as mechanical support do not interact with the guided wave and no special control of the spacing between TPP-voxels is needed. 
	These are consequently printed with a large hatching distance ($h=0.8~\mu$m).
	Finally, instead of building the entire cladding by DLW-TPP, we polymerize in a single instance the integrated photonic circuit via OPP using UV blanket irradiation. 
	Combining these three aspects we are able to reduce the printing time by $\approx$ 90$\%$ compared to classical TPP only fabrication.
	Furthermore, we show that for integrated photonic chips, where the packing density is adjusted to obtain small cross-talk, one can generally expect an acceleration of fabrication on this order.
	
	We demonstrated the high-quality of our (3+1)D \emph{flash}-TPP fabrication technique. 
	We achieved precise control over the refractive index difference $\Delta$n between waveguide core and cladding by precisely adjusting the UV exposure dose during our blanket OPP illumination, here ranging from 0 to 9000 mJ/cm$^2$.
	We fitted the waveguide's output intensities to the LP$_{01}$ mode of waveguides for variety of UV exposure doses and diameters below the cut-off condition of the second LP mode. 
	From these fits we obtain a robust characterization of a waveguide's NA as a function of the UV dosage, and our data shows an excellent agreement with the exponential trend expected for saturation processes. 
	This confirmed that we can accurately tune the refractive index of the cladding via adjusting the OPP dosage, obtaining numerical aperture values ranging between NA = 0.10 and NA = 0.16.  
	We finally determined low propagation losses of -1.36 dB/mm and a continuous characterization of our 3D-printed waveguides spanning 120 days demonstrated their temporal stability.
	In DLW of waveguides in fused silica, post fabrication thermal annealing reduced losses from similar levels down to $\sim0.08~$dB/cm, which could be a promising avenue for approaching the material's absorption limit.
	
	Overall, we have established the (3+1)D \emph{flash}-TPP proof of concept as a powerful manufacturing technique towards the fabrication of polymer-based 3D integrated photonic circuits.
	Our new concept can be particularly advantageous in applications requiring integrated photonics circuits of complex 3D topology that render a standard 2D integration impossible. 
	On this line, our approach presents a very promising concept for future scalable integrations of parallel optical interconnects.

	\section*{Funding}
	European Union’s Horizon 2020 (713694); Volkswagen Foundation (NeuroQNet II); Agence Nationale de la Recherche (ANR-17-EURE-0002, ANR-15-IDEX-0003); Region Bourgogne Franche-Comté; French Investissements d’Avenir; French RENATECH network and FEMTO-ST technological facility.
	
\bibliography{references}
\bibliographystyle{ieeetr}
	
\end{document}